\documentstyle[11pt,rotate,cite,epsf]{article}
\begin{document}
\title{
\LARGE{\bf Is $f_1(1420)$ the partner of $f_1(1285)$ 
in the $^3P_1$ $q\bar{q}$ nonet ?}\footnote{Supported in 
part by the National Natural Science
Foundation of China under Grant No. 19991487, and Grant No. LWTZ-1298
of the Chinese Academy of Sciences.}}
\vspace{1cm} 
\author{\small  De-Min Li, Hong Yu and Qi-Xing Shen\\ 
\small Institute of High Energy Physics, Chinese Academy of Sciences,\\
\small P.O.Box 918 (4), Beijing $100039$, China}
\date{}
\maketitle

\vspace*{0.3cm}
\begin{abstract}

Based on a $2\times 2$ mass matrix, the mixing angle of the axial vector
states
$f_1(1420)$ and $f_1(1285)$ is determined to be $51.5^{\circ}$, and the
theoretical 
results about the decay and production of the two states are presented.  
The theoretical results are in good 
agreement with the present experimental results, which suggests that   
$f_1(1420)$ can be assigned as the partner of $f_1(1285)$ in
the $^3P_1$ $q\bar{q}$ nonet. We also suggest that the existence of
$f_1(1510)$ needs further experimental confirmation.
\end{abstract}
\vspace{1.5cm}


\newpage

\baselineskip 24pt
\indent

The quark model predicts that there are two isoscalar states in the 
$^{3}P_1$ $q\bar{q}$ nonet, but to date there are three states 
$f_1(1285)$, $f_1(1420)$ and $f_1(1510)$ with $I=0$ and $J^{PC}=1^{++}$  
listed by Particle Data Group (PDG)\cite{1}. For $f_1(1285)$, it is 
believed that it
is well established as the $u\bar{u}+d\bar{d}$ member of the $^{3}P_1$
$q\bar{q}$ nonet. Therefore, $f_1(1420)$ and $f_1(1510)$ compete for
the $s\bar{s}$ assignment in the $^{3}P_1$ $q\bar{q}$ nonet, and one of
them must be a non-$q\bar{q}$ state.  
On one hand, in $K^{-}p~\rightarrow ~\Lambda K\bar{K}\pi$\cite{2,3} 
$f_1(1510)$ has been observed but not $f_1(1420)$. 
On the
other hand, $f_1(1420)$ has been reported in $K^{-}p$ but not in
$\pi^{-}p$\cite{4}, while two experiments do not observe $f_1(1510)$ in
$K^{-}p$\cite{4,5}. These facts make the classification of $f_1(1420)$
and $f_1(1510)$ controversial\cite{2,6,7,8,9,10}. However, the
absence of $f_1(1510)$ in radiative
$J/\psi$\cite{11,12}, central collisions\cite{13} and $\gamma\gamma$
collisions\cite{14} leads to the conclusion that $f_1(1510)$ seems not 
to be well established and its assignment as the $s\bar{s}$ member of 
the $^3P_1$ $q\bar{q}$ nonet is premature\cite{10}. 

Recently, F.E. Close {\sl et al} assigned  $f_1(1420)$ as the 
partner of $f_1(1285)$ by applying their glueball-$q\bar{q}$
filter technique to the axial
vector nonet\cite{10}. In this letter, we shall discuss the possibility 
of $f_1(1420)$ being the partner of $f_1(1285)$ in the $^3P_1$
$q\bar{q}$ nonet by studying the 
mixing effects of $f_1(1420)$ and $f_1(1285)$.

In the $|S\rangle=|s\bar{s}\rangle$,
$|N\rangle=|u\bar{u}+d\bar{d}\rangle/\sqrt{2}$ basis, the 
mass square matrix describing the quarkonia-quarkonia mixing can be 
written as follows\cite{15}:
\begin{equation}
M^2=\left (\begin{array}{cc}
M^2_S+A_S & \sqrt{2}A_{SN}\\
\sqrt{2}A_{NS} & M^2_N+2A_N
\end{array}\right),
\end{equation}
where $M_S$ and $M_N$ are the masses of the bare states $|S\rangle$ and
$|N\rangle$, 
respectively; $A_S$, $A_N$, $A_{SN}$ and $A_{NS}$ are the mixing 
parameters which describe the quarkonia-quarkonia transition amplitudes. 
Here, we assume that the physical states $|f_1(1420)\rangle$ and
$|f_1(1285)\rangle$ 
are the eigenstates of the matrix $M^2$ with the eigenvalues of $M^2_1$ and 
$M^2_2$, respectively. From the above we can get the following equations:
\begin{equation}
M^2_S+M^2_N+2A_N+A_S=M^2_1+M^2_1,
\end{equation}
\begin{equation}
(M^2_S+A_S)(M^2_N+2A_N)-2A_{SN}A_{NS}=M^2_1M^2_2.
\end{equation}
According to the factorization hypothesis\cite{15}
\begin{equation}
A_{SN}=A_{NS}\equiv\sqrt{A_NA_S},
\end{equation}
we can introduce a parameter $R$ to get
\begin{equation}
A_{NS}=A_NR,~~A_S=A_NR^2,
\end{equation}
with $0<R\leq 1$.
From Eqs. $(2)$, $(3)$ and $(5)$, $A_N$ and $R$
can be expressed as
\begin{eqnarray}
A_N=\frac{(M^2_N-M^2_1)(M^2_N-M^2_2)}{2(M^2_S-M^2_N)},\\
R^2=\frac{2(M^2_1-M^2_S)(M^2_S-M^2_2)}{(M^2_N-M^2_1)(M^2_N-M^2_2)}.
\end{eqnarray}
Diagonalizing the matrix $M^2$, we can get a unitary matrix $U$ which 
transforms the states $|S\rangle$ and $|N\rangle$ to the physical states 
$|f_1(1420)\rangle$ and $|f_1(1285)\rangle$,
\begin{equation}
U=\left( \begin{array}{cc}
x_1&y_1\\
x_2&y_2
\end{array}\right )
=\left( \begin{array}{cc}
\frac{\sqrt{2}A_NR}{-C_1} &\frac{M^2_1-M^2_S-A_NR^2}{-C_1}\\
\frac{\sqrt{2}A_NR}{-C_2} &\frac{M^2_2-M^2_S-A_NR^2}{-C_2}  
\end{array}\right)
\end{equation}
with $C_1=\sqrt{2A^2_NR^2+(M^2_1-M^2_S-A_NR^2)^2}$, 
$C_2=\sqrt{2A^2_NR^2+(M^2_2-M^2_S-A_NR^2)^2}$.

We choose $M_1=1.4262$ GeV, $M_2=1.2819$ GeV and $M_N=1.23$ GeV,
the mass of the isovector state 
$a_1(1260)$\cite{1}. $M_S$ can be obtained from the Gell-Mann-Okubo mass 
formula $M^2_S=2M^2_{K_{1A}}-M^2_{a_1}$\cite{16}, 
where $M_{K_{1A}}=1.313$ 
GeV\cite{17}. We take $M_1$, $M_2$, $M_N$ and $M_S$ as input and get the
numerical form of the matrix $U$ as follows:
\begin{equation}
U=\left(\begin{array}{cc}
x_1&y_1\\
x_2&y_2
\end{array}\right)
=\left(\begin{array}{cc}
-0.96 &-0.28\\
-0.28 &0.96
\end{array}\right).
\end{equation}
The physical states $|f_1(1420)\rangle$ and $|f_1(1285)\rangle$ can be
read as
\begin{eqnarray}
|f_1(1420)\rangle=-0.96|S\rangle-0.28|N\rangle,\\
|f_1(1285)\rangle=-0.28|S\rangle+0.96|N\rangle.
\end{eqnarray}
If we re-express the physical states $|f_1(1420)\rangle$ and
$|f_1(1285)\rangle$ in 
the Gell-Mann basis
$|8\rangle=|u\bar{u}+d\bar{d}-2s\bar{s}\rangle/\sqrt{6}$, 
$|1\rangle=|u\bar{u}+d\bar{d}+s\bar{s}\rangle/\sqrt{3}$, the 
phyical states $|f_1(1420)\rangle$ and 
$|f_1(1285)\rangle$ can be read as
\begin{eqnarray}
|f_1(1420)\rangle=\cos\theta|8\rangle-\sin\theta|1\rangle,\\
|f_1(1285)\rangle=\sin\theta|8\rangle+\cos\theta|1\rangle
\end{eqnarray}
with $\theta=51.5^{\circ}$, which is in good agreement with the result 
$\theta\sim 50^{\circ}$ given by Close {\sl et al.}\cite{10}.

Performing an elementary $SU(3)$ calculation\cite{18,19,20}, 
we obtain
\begin{equation}
\frac{Br(f_1(1285)\rightarrow\phi\gamma)}
{Br(f_1(1285)\rightarrow\rho\gamma)}=\frac{4}{9}
\left(\frac{P_{\phi}}{P_{\rho}}\right)^3
\left(\frac{x_2}{y_2}\right)^2=0.007,
\end{equation}
\begin{equation}
\frac{\Gamma(f_1(1420)\rightarrow \gamma\gamma)}
{\Gamma(f_1(1285)\rightarrow \gamma\gamma)}=
\left(\frac{M_1}{M_2}\right)^3
\frac{(y_1+\sqrt{2}x_1/5)^2}
{(y_2+\sqrt{2}x_2/5)^2}=0.539,
\end{equation}
\begin{equation}
\frac{Br(J/\psi\rightarrow \gamma f_1(1420)}
{Br(J/\psi\rightarrow \gamma f_1(1285)}=\left(\frac{P_{f_1(1420)}}
{P_{f_1(1285)}}\right)^3\frac{(\sqrt{2}y_1+x_1)^2}{(\sqrt{2}y_2+x_2)^2}
=1.36,
\end{equation}
\begin{equation}
\frac{Br(J/\psi\rightarrow f_1(1420)\omega)}
{Br(J/\psi\rightarrow f_1(1285)\phi)}=\frac{P_{\omega}}{P_{\phi}}
\left(\frac{y_1}{x_2(1-2s)}\right)^2>1.029,
\end{equation}
where $P_j$ ($j=\phi,~\rho,~\omega,~f_1(1285),~f_1(1420)$) is the 
momentum of the final state meson $j$ in the center of mass system, and $s$ 
is a parameter describing the effects of $SU(3)$ breaking. For the axial 
vector mesons, we expect $0<s\leq 1$\cite{20}. Let $r_{th}$ stand for 
the ratio of Eq. (15) to Eq. (16), so we have
\begin{equation}
r_{th}=0.397.
\end{equation}
Using the data cited by PDG\cite{1}, we have
\begin{equation}
\frac{Br(f_1(1285)\rightarrow\phi\gamma)}
{Br(f_1(1285)\rightarrow\rho\gamma)}=
\frac{(7.9\pm 3.0)\times 10^{-4}}{(5.4\pm 1.2)\times 10^{-2}}=0.015\pm 0.009,
\end{equation}
\begin{equation}
\frac{\Gamma(f_1(1420)\rightarrow \gamma\gamma)}
{\Gamma(f_1(1285)\rightarrow \gamma\gamma)}=
\frac{0.34\pm 0.18}{Br(f_1(1420)\rightarrow K\bar{K}\pi)},
\end{equation}
\begin{equation}
\frac{Br(J/\psi\rightarrow \gamma f_1(1420)}
{Br(J/\psi\rightarrow \gamma f_1(1285)}=\frac{(8.3\pm 1.5)\times 10^{-4}}
{(6.5\pm 1.0)\times 10^{-4}}\frac{1}{Br(f_1(1420)\rightarrow K\bar{K}\pi)},
\end{equation}
\begin{equation}
\frac{Br(J/\psi\rightarrow f_1(1420)\omega)}
{Br(J/\psi\rightarrow f_1(1285)\phi)}=\frac{(6.8\pm 2.4)\times 10^{-4}}  
{(2.6\pm 0.5)\times 10^{-4}}=2.62\pm 1.42.
\end{equation}
The ratio $r_{ex}$ of Eq. (20) to Eq. (21) gives 
\begin{equation}
r_{ex}=0.27\pm 0.23.
\end{equation}

From Eqs. (14)$\sim$(23), we find that the theoretical results are in good 
agreement with the experimental results, i.e., 
$|f_1(1420)\rangle=-0.96|S\rangle-0.28|N\rangle$ and
$|f_1(1285)\rangle=-0.28|S\rangle+0.96|N\rangle$ are
compatible with the experimental results. This suggests that the
present experimental results support the assignment of $f_1(1420)$ as the
partner of $f_1(1285)$ in the $^3P_1$ $q\bar{q}$ nonet.

We do not intend to discuss in detail the questionable interpretation 
of $f_1(1510)$ with dominant $s\bar{s}$ structure and that of the 
non-$q\bar{q}$ nature of $f_1(1420)$ in this letter (see ref.\cite{10}). 
However, from our results, we 
believe that if $f_1(1510)$ with $I=0$ and $J^{PC}=1^{++}$ really 
exists, it should be a non-$q\bar{q}$ state.  
Its existence of $f_1(1510)$ needs further experimental confirmation. 

In conclusion, by studying the mixing effects of $f_1(1285)$ and 
$f_1(1420)$, we find that the present experimental results support the
assignment of  
$f_1(1420)$ as the partner of $f_1(1285)$ in 
the $^3P_1$ $q\bar{q}$ nonet. We also suggest that the 
existence of $f_1(1510)$ needs further experimental confirmation.

\end{document}